\documentclass[pra,aps,twocolumn,superscriptaddress,notitlepage]{revtex4-1}
\usepackage{amssymb,amsfonts,amsmath,amstext,graphicx,hyperref}
\usepackage{tikz}
\usepackage{graphicx}
\usepackage[normalem]{ulem}
\usepackage{tabularx}

\def\BraVert{\egroup\,\mid\,\bgroup}

\usepackage{braket}
\usepackage{mathtools}
\hypersetup{
	colorlinks=true,
	linkcolor=blue,
	filecolor=magenta,      
	urlcolor=cyan,
}
\usepackage{xcolor}

\newcommand{\fplus}[1][black]{%
	\begingroup\leavevmode\color{#1}%
	\setlength{\unitlength}{0.5em}%
	\linethickness{.15em}%
	\begin{picture}(1,1)
	\put(0,0.5){\line(1,0){1}}
	\put(0.5,0){\line(0,1){1}}
	\end{picture}%
	\endgroup
}
\newcommand{\sq}[1][black]{%
	\begingroup\leavevmode\color{#1}%
	\setlength{\unitlength}{0.5em}%
	\linethickness{.5em}%
	\begin{picture}(1,1)
	\put(0,0.5){\line(1,0){1}}
	\put(0.5,0){\line(0,1){1}}
	\end{picture}%
	\endgroup
}
\newcommand{\blackdiam}{\rotatebox[origin=c]{45}{$\sq$}}

\begin{document}

	\title{Exotic synchronization in continuous time crystals outside the symmetric subspace}
	
	\author{Parvinder Solanki}
	\email{parvinder.parvinder@unibas.ch}
	\affiliation{Department of Physics, University of Basel, Klingelbergstrasse 82, CH-4056 Basel, Switzerland}
	
	\author{Midhun Krishna}
	\affiliation{Department of Physics, Indian Institute of Technology-Bombay, Powai, Mumbai 400076, India}

	\author{Michal Hajdu\v{s}ek}
	\email{michal@sfc.wide.ad.jp}
	\affiliation{Keio University Shonan Fujisawa Campus, 5322 Endo, Fujisawa, Kanagawa 252-0882, Japan}
	\affiliation{Keio University Quantum Computing Center, 3-14-1 Hiyoshi, Kohoku, Yokohama, Kanagawa 223-8522, Japan}
	\author{Christoph Bruder}
	\affiliation{Department of Physics, University of Basel, Klingelbergstrasse 82, CH-4056 Basel, Switzerland}
	\author{Sai Vinjanampathy}
	\email{sai@phy.iitb.ac.in}
	\affiliation{Department of Physics, Indian Institute of Technology-Bombay, Powai, Mumbai 400076, India}
	\affiliation{Centre of Excellence in Quantum Information, Computation, Science and Technology, Indian Institute of Technology Bombay, Powai, Mumbai 400076, India}
	\affiliation{Centre for Quantum Technologies, National University of Singapore, 3 Science Drive 2, Singapore 117543}
	\date{\today}

	\begin{abstract}
		Exploring continuous time crystals (CTCs) within the symmetric subspace of spin systems has been a subject of intensive research in recent times. Thus far, the stability of the time-crystal phase outside the symmetric subspace in such spin systems has gone largely unexplored. Here, we investigate the effect of including the asymmetric subspaces on the dynamics of CTCs in a driven dissipative spin model. This results in multistability, and the dynamics becomes dependent on the initial state.  Remarkably, this multistability leads to exotic synchronization regimes such as chimera states and cluster synchronization in an ensemble of coupled identical CTCs. Interestingly, it leads to other nonlinear phenomena such as oscillation death and signature of chaos.
		
	\end{abstract} 
	\maketitle

	\noindent \textit{Introduction.---} Crystals have traditionally been characterized by their spatial periodicity, wherein atoms are arranged in a repeating pattern. 
	Time crystals showcase a distinct form of periodicity, namely in the temporal dimension~\cite{wilczek2012quantum,watanabe2015absence,RevModPhys.95.031001,sacha2017time}.
	Initially, time crystals breaking discrete time-translation symmetry were conceptualized and termed discrete time crystals \cite{sacha2015modeling,else2016floquet,khemani2016phase,russomanno2017floquet,surace2017floquet,heugel2019classical,khasseh2019many,sakurai2021chimera, sacha2017time, RevModPhys.95.031001}. 
	A more recent development in the field has been the emergence of continuous time crystals (CTCs) \cite{iemini2018boundary, tucker2018shattered,buca2019non,Booker_2020,PhysRevLett.123.260401,zhu2019dicke,lledo2019driven,seibold2020dissipative,prazeres2021boundary,piccitto2021symmetries,carollo2022exact,seeding2022michal,krishna2022measurement,hurtado2020raretimecrystal,PhysRevB.108.024302,PhysRevLett.127.133601,cabot2023nonequilibrium}, which break continuous time-translation symmetry. 
	CTCs were initially discussed in the context of a driven dissipative Dicke model \cite{iemini2018boundary}. 
	Their behavior was also studied in other spin systems like collective $d$-level systems \cite{prazeres2021boundary}, $p$-$q$ interacting spin models \cite{piccitto2021symmetries}, spin star models \cite{krishna2022measurement}, and spin systems with infinite-range interactions \cite{PhysRevB.108.024302}. 
	In addition to the time-crystal behavior, the thermodynamics \cite{igor_thermodynamics} of driven Dicke models and the occurrence of entanglement phase transitions \cite{saro_postselection, albert_trajectory} have also been investigated recently.
	The application of time-crystal phases in sensing \cite{cabot2023continuous,montenegro2023quantum} and quantum engines \cite{carollo2020engines} has been proposed.
	Furthermore, experimental platforms such as Bose-Einstein condensate systems have been used to realize CTCs \cite{kessler2021observation,kongkhambut2022observation}.

	All of these examples are variants of the Dicke model: the system is a collective spin that is built up of $N$ spins 1/2. These systems are characterized by permutational symmetry which is a strong symmetry of the underlying open quantum system~\cite{buca2012,albert2014symmetries}.
	Previous investigations of both superradiance \cite{dicke1954coherence,AnatoliiVAndreev_1980,wang1973phase,scully2009super} and time-crystal behavior have focused on the totally symmetric Dicke subspace, which corresponds to the maximum-excitation subspace.
	The assumption of permutational symmetry is not always straightforward to enforce under realistic conditions \cite{ferioli2023non}. 
	This leads to the important open question of the role played by the lower-excitation subspaces involving asymmetric Dicke states in the dynamics of time crystals. Recent studies have shown that asymmetric subspaces are crucial for understanding subradiant phases~\cite{PhysRevLett.120.113603,PhysRevA.106.053712, gegg2018superradiant,PhysRevA.84.023805} and the thermodynamics of permutation-invariant quantum many-body systems \cite{latune2019thermodynamics,Latune_2020,yadin2023thermo}.

	In this Letter, we investigate the existence of time crystals outside the symmetric Dicke subspace in a spin-only description of the driven dissipative Dicke model \cite{iemini2018boundary,seeding2022michal}.
	We show that incorporating the asymmetric subspaces enhances the complexity and richness of CTC dynamics and yields an extended phase diagram.
	Using mean-field analysis, we find that CTCs exhibit multistability in the asymmetric subspaces, which was not observed for CTCs restricted to the symmetric subspace.
	The existence of multistability is confirmed by analyzing the eigenspectra of the Liouville superoperator that describes the Markovian dynamics of open quantum systems \cite{buca2012,victor2014,krishna2023select}. 
	Depending on the system parameters, an initial state outside the symmetric subspace can always be found which breaks time-translational symmetry. 
	Interestingly, this initial-state dependence leads to exotic synchronization phenomena in a network of coupled identical CTCs 
	like chimera states that are characterized by the coexistence of synchronized and unsynchronized regions~\cite{kuramoto2002coexistence,Abrams2004,multistability1,multistability2,multistability3}.
	Furthermore, such a network of CTCs also exhibits cluster synchronization, where identical oscillators arrange themselves into two or more differently synchronized clusters~\cite{Haugland_2021}.
	Finally, we show that this highly nonlinear coupled system also features chaotic mean-field dynamics and the phenomenon of oscillation death.
	
	\begin{figure}[htp!]
		\centering    \includegraphics[width=\linewidth]{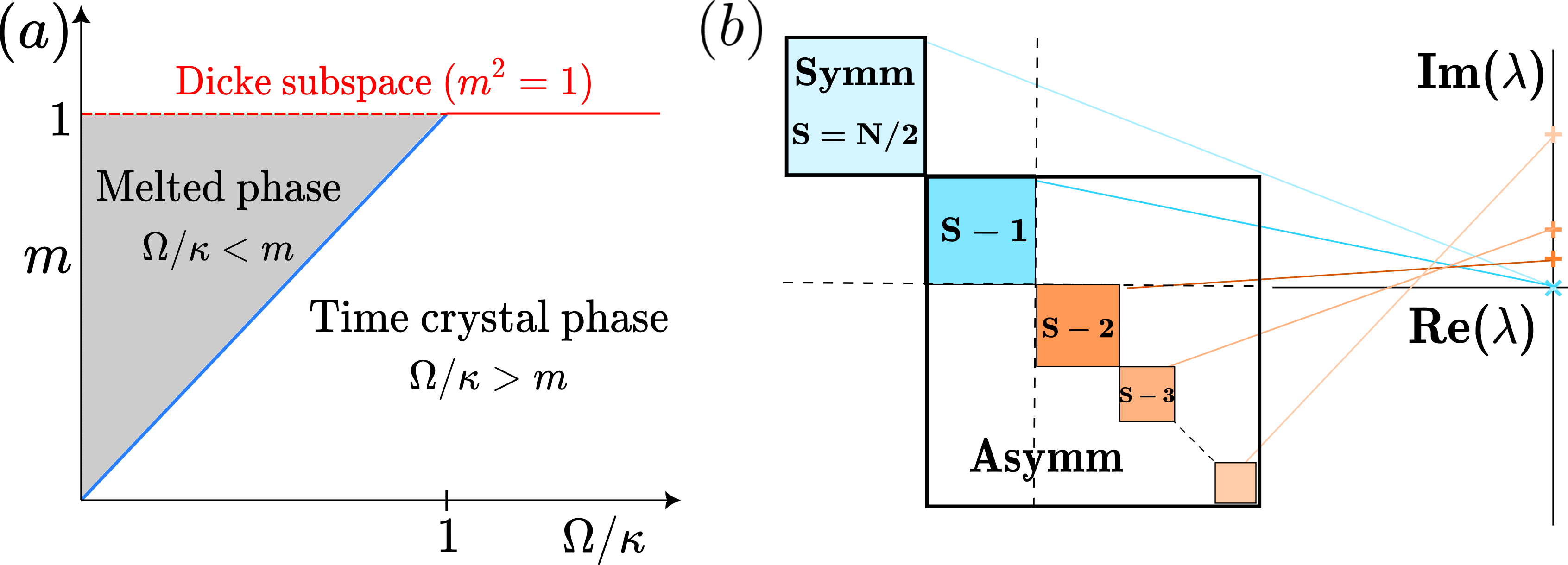}
		\caption
		{(a) 
			Phase diagram of a continuous time crystal (CTC) characterized using fixed-point analysis of mean-field equations with $m=\langle \hat{S} \rangle/S$. (b) Schematic representation of block diagonal structure of Liouville superoperator $\mathcal{L}=\mathcal{L}_S+\mathcal{L}_A$ where $\mathcal{L}_{S(A)}$ corresponds to symmetric (asymmetric) Dicke subspace. Here, dotted lines separate the blocks with $m>\Omega/\kappa$ from those with $m<\Omega/\kappa$.  Only the sub-blocks having rescaled magnetization value $m<\Omega/\kappa$ will have purely imaginary eigenvalues in the thermodynamic limit $N=2S\rightarrow \infty$. } 
		\label{fig:intro}
	\end{figure}
	
	We begin with an investigation of the mean-field dynamics of a collective spin using fixed-point analysis, which describes the steady-state dynamics of the system.
	Then, we will use Liouville superoperator theory to validate the mean-field results.
	Finally, we discuss the existence of a chimera state, cluster synchronization, oscillation death and chaos in coupled CTCs before summarizing our results.

	\noindent \textit{Mean-field analysis and phase diagram.---} The master equation governing the dynamics of the CTC \cite{iemini2018boundary,seeding2022michal} is given by
	\begin{equation}\label{eq:master_BTC}
	\dot{\rho }=\mathcal{L}_{\Omega,\kappa }(\rho)=-i[\Omega \hat{S}_x ,\rho]+\frac{\kappa}{S}(\hat{S}_-\rho\hat{S}_+ -\frac{1}{2}\{\hat{S}_+ \hat{S}_-,\rho \}),
	\end{equation}
	where $\mathcal{L}_{\Omega ,\kappa }$ is the Liouville superoperator, $\hat{S}_{\alpha }=\sum_i \sigma^i_{\alpha }/2 $ are collective spin operators where $\alpha\in \{ x,y,z \} $, $S=N/2$ is the maximum spin value for $N$ spins $1/2$ and $\hat{S}_{\pm }=\hat{S }_x \pm i\hat{S }_y$. 
	The dynamics of the system in the thermodynamic limit ($N\rightarrow \infty$) can be well understood using mean-field analysis. 
	The evolution of mean-field expectation values of the rescaled spin operators, $m_{\alpha }=\langle \hat{S}_{\alpha }\rangle /S $ is governed by
	\begin{align} \label{eq:btcmf123}
	\dot{m}_x&=\kappa m_x m_z, \nonumber\\ 
	\dot{m}_y&= -\Omega m_z+ \kappa m_y m_z, \\ 
	\dot{m}_z&= \Omega m_y- \kappa [m_x^2+m_y^2].\nonumber
	\end{align}
	The total spin of the system is conserved since $[\hat{S}^2,S_{x,y,z,\pm}]=0$ 
	and hence is a strong symmetry of the system \cite{buca2012,victor2014,manzano2018harnessing}.
	At the mean-field level, this implies that $m^2_x +m^2_y +m^2_z =m^2 $ is a conserved quantity since correlations are assumed to vanish \cite{CTC_Correlation}.

	Using the fixed-point analysis, we can characterize the phase transition of the system as described by Fig.~\ref{fig:intro}(a) (see \cite{supp} for details). 
	Initial states on the left of the $m=\Omega/\kappa$ (blue) line ($m > \Omega/\kappa$) settle down to a saddle fixed point. 
	In the parameter region $m < \Omega/\kappa$, the systems exhibit stable oscillations around the center fixed point leading to 
	time-translational symmetry breaking.
	The frequency of oscillations is determined by the imaginary eigenvalues of the Jacobian matrix $\pm i \kappa \sqrt{(\Omega/\kappa)^2-m^2}$.
	The symmetric Dicke subspace corresponds to $m=1$, which was shown to lead to the breaking of time-translational symmetry at $\Omega/\kappa=1$ in earlier studies \cite{iemini2018boundary,seeding2022michal}.
	The system exhibits multistability for $\Omega/\kappa<1$. 
	Depending upon the initial state, the system can exhibit two different phases, namely, a melted and a time-crystal phase.

	\noindent \textit{Role of asymmetric subspaces.---}
	We now go beyond mean-field theory and study the full master equation Eq.~(\ref{eq:master_BTC}).
	The Hilbert space of the ensemble consisting of $N$ spins can be described as a direct sum of irreducible representations of $SU(2)$ using the Clebsch-Gordan decomposition \cite{sakurai1995modern}.
	Since $\hat{S}^2$ is a strong symmetry of the dynamics, the Liouville superoperator block-diagonalizes \cite{buca2012,thingna2021degenerate} in respective spin-$J$ eigenbases such that $\mathcal{L}=\oplus \mathcal{L}_{J}$, as shown in Fig.~\ref{fig:intro}(b). 
	Therefore, the dynamics is constrained within each spin-$J$ subspace (invariant subspaces of dynamics).
	The maximum angular momentum value $J=S=N/2$ corresponds to the totally symmetric Dicke subspace; lower $J$ values correspond to asymmetric subspaces. The spin-$J$ irreducible representation has dimension $2J+1$ and multiplicity $n_{J} = N! (2J+1)/(N/2+J+1)!(N/2-J)!$~\cite{georgi2000lie,ramadevi2019group,yadin2023thermo}.
	The total Hilbert space is spanned by the generalized Dicke basis, $\{ |J,m_{J},\mu_J \rangle \}$.
	Here, $J \in \{S,S-1 \ldots J_{0}\}$, $m_J \in \{J,J-1 \ldots -(J-1),-J\}$ and $ 1 \leq \mu_J \leq n_J$, with $J_0 = 0 \text{ or } 1$ depending on whether $N$ is even or odd.
	We can associate each of the spin-$J$ subspaces with the mean-field total angular momentum value $m=J/S$.
	The states with $m<1$ belong to the asymmetric angular momentum sectors.
	
	\begin{figure}[htp!]
		\centering    
		\includegraphics[width=\linewidth]{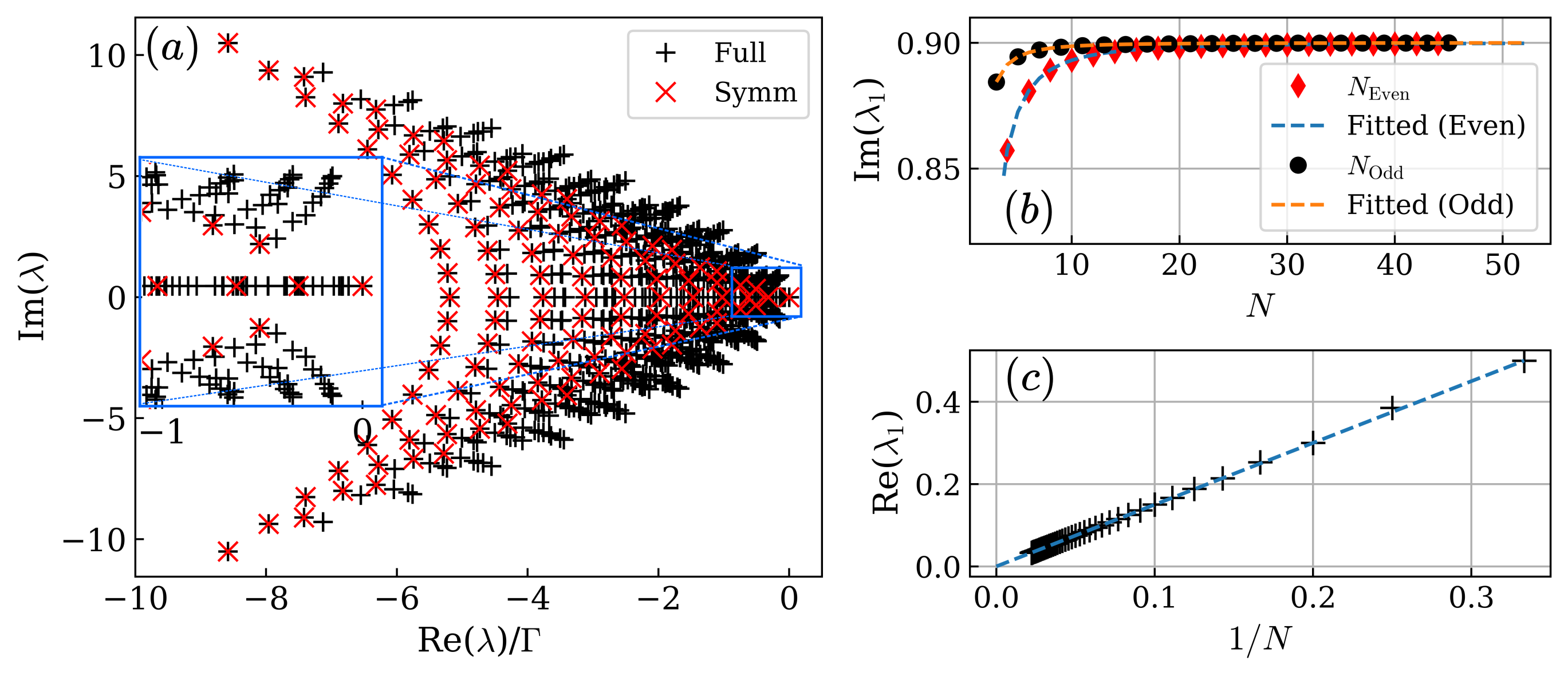}
		\caption
		{(a) Eigenspectra of $\mathcal{L}_{\Omega,\kappa}$ for symmetric and full Hilbert space (indicated by `$\times$' and `+', respectively) with $\Omega/\kappa =0.9$ (same for all subfigures) and $N=10$.
			(b) The imaginary part of the first dominant eigenvalue 
			$\lambda_1$ tends to $0.9$ (i.e., the value of $\Omega$) in the thermodynamic limit 
			and is consistent with mean-field analysis.
			(c) The real part of $\lambda_1$ scales as $1/N$ and tends to zero in the thermodynamic limit. All the eigenvalues shown here are in units of $\kappa$.
		} 
		\label{fig:eigL}
	\end{figure}
	
	\begin{figure*}[htp!]
		\centering    
		\includegraphics[width=\textwidth]{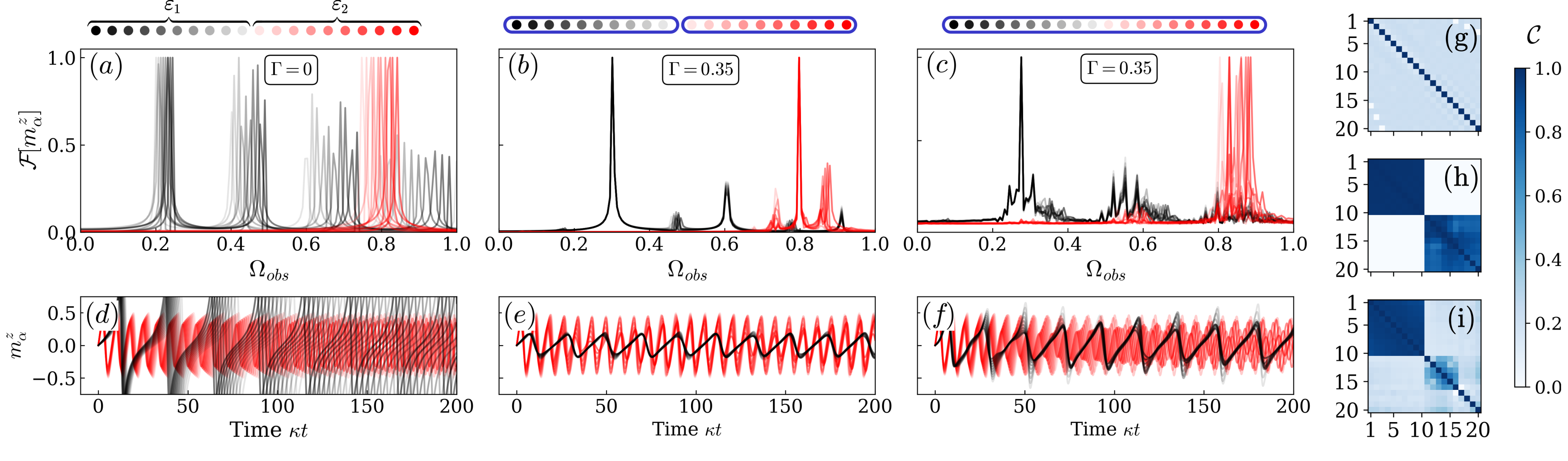}
		\caption{Normalized Fourier transform $\mathcal{F}[m^z_\alpha]$ (a-c), the time evolution of $m^z_\alpha$ (d-f) and Pearson’s correlation matrix (g-i) for $n=20$ continuous time crystals (CTCs). All the CTCs are identical ($\Omega_\alpha/\kappa=0.9$ and $\kappa=1$) and are initialized in different $m$ subspaces. The two ensembles $\varepsilon_{1,2}$ are represented by the colors black ($n_{\varepsilon_1} \coloneqq \{1,2,\ldots,10\}$) and red ($n_{\varepsilon_2} \coloneqq \{11,12,\ldots,20\}$)  above subfigures (a,b,c), and the CTCs within one ensemble by different intensities. The frequencies of the CTCs are uniformly sampled in the window from $0.2$ to $0.25$ in ensemble $\varepsilon_1$  and from $0.75$ to $0.85$ in ensemble $\varepsilon_2$. Subfigures (a,d,g) show the case of uncoupled time crystals. Coupling CTCs within each ensemble with a coupling strength of $\Gamma/\kappa=0.35$ results in the synchronization of CTCs within the ensemble, as shown in (b,e,h). 
			In subfigures (c,f,i), all CTCs are coupled. Surprisingly, CTCs in ensemble $\varepsilon_1$ get synchronized, while the CTCs in the ensemble $\varepsilon_2$ remain unsynchronized: the system exhibits a \textit{chimera} state. }
		\label{fig:chimera}
	\end{figure*}

	The time-crystal phase is characterized by the spectral gap closing of the Liouvillian eigenspectra in the thermodynamic limit, with the leading eigenvalue being purely imaginary.  
	For the symmetric subspace, the spectral gap of $\mathcal{L}_S$ closes only for $\Omega/\kappa>1$.
	Including the asymmetric subspaces $\mathcal{L}_A=\oplus_{J<S} \mathcal{L}_J$ introduces new leading-order eigenvalues, as shown in Fig.~\ref{fig:eigL}(a). Therefore, the spectral gap closes even for $\Omega/\kappa <1$, as depicted in Figs.~\ref{fig:eigL}(b)--(c). 
	The first dominant eigenvalue in the asymmetric subspace scales differently for even and odd $N$, but converges to the same value for large values of $N$, as shown in Fig.~\ref{fig:eigL}(b). 
	The real part of the eigenvalue $\lambda_1$ for both even and odd values of $N$ scales as $1/N$ with the same slope as shown in Fig.~\ref{fig:eigL}(c) and vanishes in the thermodynamic limit.
	This means that the spectral gap closes even for $\Omega/\kappa<1$ for the asymmetric subspaces.
	Therefore, time crystal behavior can be observed outside the permutational symmetric subspace even for $\Omega/\kappa<1$ if the initial state belongs to an asymmetric subspace having $m<\Omega/\kappa$.
	
	\begin{figure*}[htp!]
		\centering    \includegraphics[width=\linewidth]{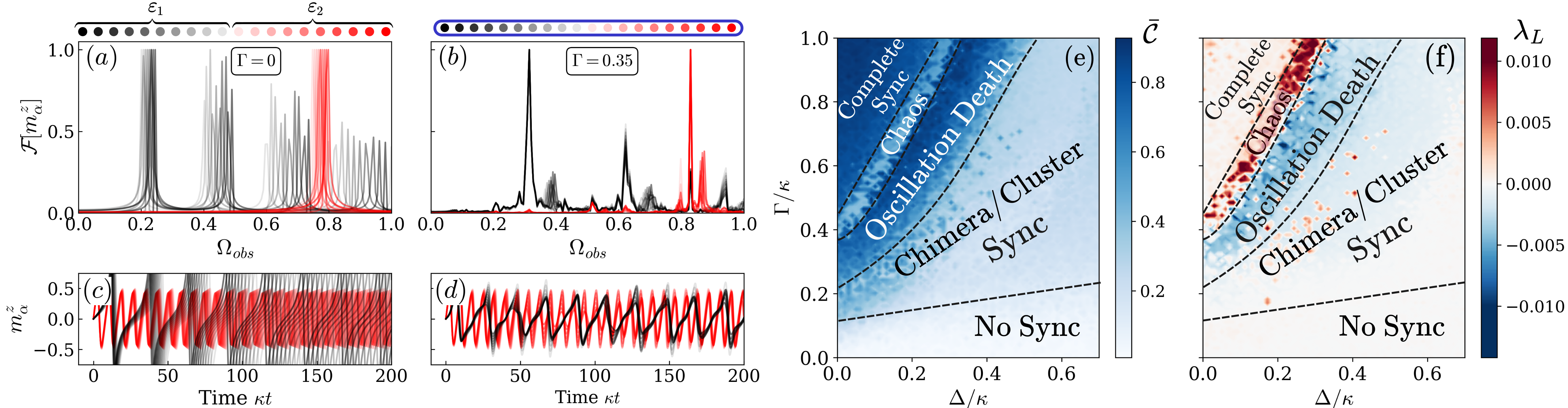}
		\caption
		{Subfigures (a-d) represent the cluster synchronization where (a,c) describe the dynamics of uncoupled CTCs ($\Gamma=0$), which get synchronized in different clusters for $\Gamma=0.35$ as shown in (b,d). The frequencies of the uncoupled CTCs in (a) have a uniform distribution from $0.2$ to $0.25$ in ensemble $\varepsilon_1$  and from $0.75$ to $0.8$ in ensemble $\varepsilon_2$. Various exotic synchronization regimes are characterized by investigating (e) the average value of the Pearson coefficient and (f) the maximum Lyapunov exponent. They are studied as a function of coupling strength $\Gamma$ and detuning $\Delta$ between the mean frequencies of two Gaussian-distributed ensembles $\varepsilon_{1,2}$, for $n=100$ CTCs with standard deviations $\sigma_1=\sigma_2=0.1$. Here $\Omega_\alpha/\kappa=0.9$ and $\kappa=1$ for all subfigures. 
		}\label{fig:cluster_sync}
	\end{figure*}
	\noindent \textit{Chimera state and cluster synchronization.---} We now discuss the effect of including asymmetric subspaces on the dynamics of $n$ coherently coupled CTCs described by the following master equation,
	\begin{equation}
	\dot \rho = \sum_\alpha^n \mathcal{L}_{\Omega_\alpha,\kappa_\alpha} (\rho) -\frac{i}{n}[\sum_{\beta \neq \alpha}^n \Gamma_{\alpha,\beta}(\hat{S}^\alpha_+\hat{S}^\beta_- +\hat{S}^\alpha_- \hat{S}^\beta_+),\rho].\label{eq:coupledCTC}
	\end{equation}
	Here, $\mathcal{L}_{\Omega_\alpha,\kappa_\alpha}(\rho)$ describes the dynamics of uncoupled CTCs, and $\Gamma_{\alpha,\beta}$ is the coupling strength between two CTCs $\alpha$ and $\beta$. 
	Such an interaction resulted in the observation of the seeding effect and the synchronization of detuned CTCs considering only the symmetric subspace~\cite{seeding2022michal}. 
	We now also take into account asymmetric subspaces and assume that all the CTCs are identical, $\mathcal{L}_{\Omega_\alpha,\kappa_\alpha} = \mathcal{L}_{\Omega,\kappa}, ~\forall \alpha$. 
	Even if they are identical, different choices of initial states can lead to surprising disparate effects in the dynamics.
	We investigate the mean-field dynamics of $n=20$ coupled CTCs in Fig.~\ref{fig:chimera}, where all CTCs are initialized in asymmetric subspaces belonging to different values of $m$. 
	The corresponding frequency and the dynamics of $m_z^\alpha$ of uncoupled CTCs depending on the value $m$ are shown in Fig.~\ref{fig:chimera}(a) and Fig.~\ref{fig:chimera}(d), respectively.
	For simplicity, we consider two subgroups based on the separation between the frequency distribution of uncoupled CTCs, though a general distribution could result in more than two subgroups. 
	In Fig.~\ref{fig:chimera}(a), the system is split into two ensembles, ${\varepsilon_1}$ (black) and ${\varepsilon_2}$ (red), with uniform frequency distributions: ${\varepsilon_1}$ ranges from 0.2 to 0.25, and ${\varepsilon_2}$ from 0.75 to 0.85.

	We use Pearson’s correlation coefficient as an indicator of synchronization. For any two coupled CTCs $\alpha$ and $\beta$, it can be defined as follows
	\begin{equation}
	\mathcal{C}_{\alpha \beta} = \frac{\langle{m^{z}_{\alpha} m^{z}_{\beta}}\rangle- \langle{m^{z}_{\alpha}}\rangle \langle{m^{z}_{\beta}}\rangle}
	{\sqrt{(\langle{m^{z}_{\alpha})^{2}}\rangle- \langle{m^{z}_{\alpha}}\rangle ^{2}) (\langle{(m^{z}_{\beta})^{2}}\rangle- \langle{m^{z}_{\beta}}\rangle ^{2})}},
	\end{equation}
	where $\langle f \rangle = \frac{1}{T}\int_0^{T}f(t)dt$ represents the time average.
	The off-diagonal elements $\mathcal{C}_{\alpha\neq\beta}$ of the correlation matrix $\mathcal{C}_{n \times n}$ indicates the amount of correlation between the CTCs $\alpha$ and $\beta$.
	Therefore, $\mathcal{C}_{\alpha,\beta}\rightarrow 1$ corresponds to complete synchronization and $\mathcal{C}_{\alpha,\beta}\rightarrow 0$  indicates no synchronization. E.g., Fig.~\ref{fig:chimera}(g) shows that this matrix is diagonal for uncoupled CTCs.

	First, we consider the case where CTCs within the same ensemble are coupled with a finite interaction strength but there is no coupling between CTCs of different ensembles such that $\Gamma_{\alpha,\beta}=\Gamma \delta_{\varepsilon_i \varepsilon_j} ~\forall~( \alpha \in \varepsilon_i, \beta \in \varepsilon_j)$.
	This results in the synchronization of CTCs within the same ensemble as shown by Fig.~\ref{fig:chimera}(b,e,h).
	Surprisingly, considering all-to-all coupled CTCs ($\Gamma_{\alpha,\beta}/\kappa=\Gamma/\kappa=0.35~ \forall ~(\alpha,\beta)$) leads to the synchronization of the CTCs only in one ensemble (here $\varepsilon_1$), while the CTCs in the other ensemble (here $\varepsilon_2$) exhibit unsynchronized dynamics as shown in Fig.~\ref{fig:chimera}(c,f,i).
	Such surprising patterns of partial synchronization in systems of coupled identical nonlinear oscillators were first observed in~\cite{kuramoto2002coexistence} and were called \textit{chimera states}~\cite{Abrams2004,zakharova_book}. 
	First strides towards chimera states in quantum systems involved exploring one-dimensional arrays of quantum oscillators \cite{bastidas2015chimera} and periodically driven networks of spins~\cite{sakurai2021chimera}.
	Our work demonstrates a chimera state in a system of coupled identical CTCs.
	Here, the chimera state results from including the asymmetric subspaces into the dynamics of identical CTCs, allowing us to choose an initial state with different frequencies in different subspaces~\cite{ujjwal2017chimera}.
	Increasing the coupling strength results in the synchronization of all CTCs, see~\cite{supp}.
	
	Another example of an exotic synchronization regime that can be realized in coupled identical CTCs is \textit{cluster synchronization}, where the system separates into two differently synchronized clusters. Such a state is shown in Fig.~\ref{fig:cluster_sync}(b), where the group $\varepsilon_1$ (black) and the group $\varepsilon_2$ (red) of CTCs get synchronized separately at different frequencies. 
	This results from the choice of the initial state shown in Fig.~\ref{fig:cluster_sync}(a), where the frequency distribution of the ensemble $\varepsilon_1$ is the same as in Fig.~\ref{fig:chimera}(a), while it ranges from 0.75 to 0.8 for $\varepsilon_2$.

	\noindent \textit{Characterization of different phases.---}
	We now extend our analysis to investigate the effect of choosing different initial states and coupling strengths $\Gamma$ on the synchronization properties of $n$ coupled CTCs. 
	These CTCs are initialized in different $m$ subspaces such that the frequency of CTCs in each subgroup $\varepsilon_{1,2}$ are sampled from Gaussian distributions with distinct mean frequencies $\bar{\omega}_{1,2}$ and standard deviations $\sigma_{1,2}$.
	We investigate the synchronization and dynamical properties as a function of coupling strength $\Gamma$ and detuning $\Delta=\bar{\omega}_2-\bar{\omega}_1$ for fixed values of $\sigma_{1,2}$, where $\Delta$ and  $\sigma_{1,2}$ depend on the choice of initial states. 
	The mean value of Pearson’s correlation $\bar{\mathcal{C}}=\sum_{\alpha < \beta } \mathcal{C}_{\alpha \beta }/\mathcal{N}$ will be used to quantify the amount of synchronization, where $\mathcal{N}=n(n-1)/2$. 
	This measure takes the value $\bar{\mathcal{C}}=0$ for no synchronization, $\bar{\mathcal{C}}=1$ for complete synchronization, and $0<\bar{\mathcal{C}}<1$ in the regime of partial synchronization, such as the chimera and cluster-synchronized states.
	
	The synchronization measure alone is not sufficient to understand the complex Arnold-tongue structure observed in Fig.~\ref{fig:cluster_sync}(e). 
	Therefore, we also investigate the largest Lyapunov exponent $\lambda_L$ (Fig.~\ref{fig:cluster_sync}(f)), which measures the sensitivity of the dynamics of a nonlinear system to its initial conditions \cite{NLD1, NLD2}.
	A positive value of $\lambda_L$ corresponds to chaotic dynamics where two nearby trajectories diverge exponentially over time,
	whereas the system settles down to a stable fixed point for $\lambda_L<0$.
	Finally, $\lambda_L \approx 0$ represents limit cycles and quasi-periodic dynamics. 
	
	In Figs.~\ref{fig:cluster_sync}(e,f), we characterize different regimes making use of both the synchronization measure $\bar{\mathcal{C}}$ and the maximum Lyapunov exponent $\lambda_L$.
	For a finite detuning $\Delta$ and small coupling strength $\Gamma$, coupled CTCs exhibit unsynchronized dynamics indicated by $\bar{\mathcal{C}}\approx\lambda_L\approx 0$. 
	Increasing $\Gamma$ leads to a regime of partial synchronization with $\bar{\mathcal{C}}>0$ and $\lambda_L\approx 0$, which includes chimera and cluster synchronization states. 
	These two states can be distinguished by the correlation matrix $\mathcal{C}$ (see \cite{supp} for more details). 
	Further increasing $\Gamma$ leads to $\lambda_L<0$ and $\bar{\mathcal{C}}\approx 1$, causing the system to settle down into a highly-correlated fixed point. 
	This sudden termination of oscillations is termed oscillation death. Higher values of $\Gamma$ lead to a less correlated chaotic mean-field dynamics, which is characterized by $\lambda_L>0$ and $\bar{\mathcal{C}}< 1$. 
	The reduced synchronization measure results from the chaotic nature of the dynamics which makes the system weakly correlated, see Fig.~\ref{fig:cluster_sync}(e). 
	Finally, even larger values of the coupling strength lead to complete synchronization such that $\bar{\mathcal{C}}=1$. 
	Thus, depending on the parameters and initial state configuration, the system features various exotic synchronization regimes as shown in Fig.~\ref{fig:cluster_sync}(e,f).

	\noindent \textit{Conclusions.---}
	In this work, we focused on the fundamental question of whether continuous time crystals in spin models only exist if initialized in the symmetric subspace.
	Our first main result is that  
	the inclusion of asymmetric subspaces leads to time-crystal behavior even if the ratio of drive to dissipation strength  $\Omega/\kappa$ is less than one.
	A second result is that taking into account asymmetric subspaces leads to multistability: depending on the parameter values, the system can exhibit a time crystal 
	or melted phase for different initial states. 
	For an ensemble of coupled CTCs, this leads to our third result, the existence of chimera states and cluster synchronization, which is not possible for CTCs realized in the symmetric subspace.
	Finally, our system can also exhibit chaotic mean-field dynamics and oscillations death.
	Our results provide new insights into the dynamics of time-crystal phases and point out strategies in the design of quantum networks that may lead to the observation of exotic synchronization phenomena.
	
	An interesting future direction is to study the effect of weak permutational symmetry-breaking interactions, which allow the system to move between various spin sectors. 
	Such interactions are inevitable in various experimental settings and will offer further valuable insights into the fundamental characteristics of time crystals.

	\begin{acknowledgments}
		\noindent \textit{Acknowledgments.---} P.S. and S.V. thank R. Fazio for discussions.  Numerical calculations were performed using QuTiP \cite{johansson2012qutip,johansson2013qutip2} and PIQS \cite{shammah2018piqs}. M.H. was supported by [MEXT Quantum Leap Flagship Program] Grants No. JPMXS0118067285 and No. JPMXS0120319794. P.S. and C.B. acknowledge financial support from the Swiss National Science Foundation individual grant (grant no. 200020 200481). S.V. acknowledges support from DST-QUEST grant number DST/ICPS/QuST/Theme-4/2019.  While preparing the manuscript, we became aware of related work \cite{iemini2023dynamics}. 
	\end{acknowledgments}

	\newpage
	
	\appendix
	\maketitle
	\onecolumngrid
	
	\renewcommand{\thefigure}{S\arabic{figure}}
	\setcounter{figure}{0} 
	\begin{figure*}[b!]
		\centering    \includegraphics[width=0.75\textwidth]{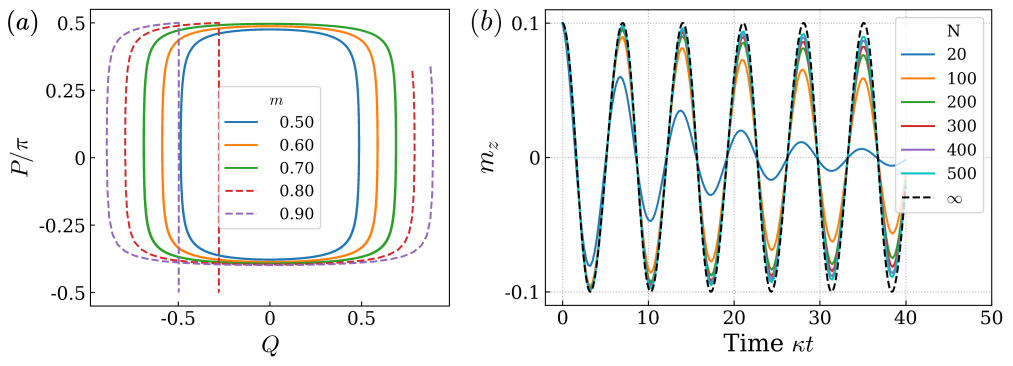}
		\caption{(a) Phase-space portrait of the CTC for various initial states, where $P=\tan^{-1}\left[ m_y/m_x \right]$ and $Q=m_z$. (b) Oscillations of $m_z$ for an initial state prepared in the asymmetric subspace $m=0.1$ for $\Omega/\kappa=0.9$. The mean-field dynamics is given by the dashed line. The time evolution of Eq.~\eqref{eq:master_BTC} of the main text for different system sizes $N$ is represented by solid lines with different colors.} 
		\label{fig:MF_TimeEvo}
	\end{figure*}

	\section{Mean-field and fixed-point analysis of a single continuous time crystal}
	In this section, we extend the mean-field analysis of continuous time crystals (CTCs) outside the symmetric subspace discussed in the main text.
	The mean-field dynamics of the system described by Eq.~\eqref{eq:master_BTC} of the main text is governed by the following set of coupled differential equations
	\begin{align} \label{eq:MF}
	\dot{m}_x&=\kappa m_x m_z, \nonumber\\ 
	\dot{m}_y&= -\Omega m_z+ \kappa m_y m_z, \\ 
	\dot{m}_z&= \Omega m_y- \kappa [m_x^2+m_y^2],\nonumber
	\end{align}
	where $m_\alpha=\langle S_\alpha \rangle/S$ are rescaled spin operators with $\alpha\in \{x,y,z\}$.
	For the symmetric Dicke subspace, $m$ is constrained to take the value $1$, which led to the breaking of time-translational symmetry at $\Omega/\kappa=1$ in earlier studies \cite{iemini2018boundary,seeding2022michal}.
	We will now consider the more general case $0 \leq m < 1$. 
	The fixed points, $(m_x^*,m_y^*,m_z^*)$, of the coupled differential Eqs.~\eqref{eq:MF} for this case are given as $\mathcal{M}_1 = (\pm m\sqrt{1-(m\kappa/\Omega)^2},m^2 \kappa/\Omega,0)$ and $\mathcal{M}_2 = (0,\Omega/\kappa,\pm \sqrt{m^2-(\Omega/\kappa)^2})$.
	The stability of these fixed points is given by the eigenvalues of the Jacobian $\mathcal{J}$, whose matrix elements are defined as $\mathcal{J}_{\alpha\beta}=d \dot{m}_\alpha/dm_\beta$.
	The eigenvalues of $\mathcal{J}$ corresponding to fixed points $\mathcal{M}_1$ and $\mathcal{M}_2$ are  $\{\Lambda_1\}=\{0,\pm \kappa \sqrt{m^2-(\Omega/\kappa)^2}\}$ and $\{\Lambda_2\}=\{0, \pm \kappa \sqrt{m^2-(\Omega/\kappa)^2},\pm \kappa \sqrt{m^2-(\Omega/\kappa)^2}\}$ respectively, where both the upper or both the lower signs have to be chosen together.
	Note that the dynamics of the system are now dependent on both the system parameters $\Omega/\kappa$ and rescaled total magnetization $m$ of the given spin-subspace. 
	For $\Omega/\kappa<m$, $\mathcal{M}_2$ is the only physical state of the system since $\mathcal{M}_1$ has unphysical imaginary values for the expectation of spin components. 
	The corresponding eigenvalues $\{\Lambda_2\}$ for the fixed point $\mathcal{M}_2$ are purely real, and hence it is a  saddle fixed point of the system. 
	Therefore, the system relaxes to a time-invariant state given by the fixed point $\mathcal{M}_2$ for initial states having $m>\Omega/\kappa$, and no time-crystal behavior is observed.
	For states with $m<\Omega/\kappa$, $\mathcal{M}_1$ is the only physical fixed point of the system.  The eigenvalues $\{\Lambda_1\}$ for the fixed point $\mathcal{M}_1$ are purely imaginary, and hence, the corresponding fixed point is a center. 
	Therefore, the system oscillates in time for initial states having $m<\Omega/\kappa$, and time-translational symmetry is broken even for $\Omega/\kappa<1$. 
	
	Using this fixed-point analysis, we can characterize the phase transition of the system as described by Fig.~\ref{fig:intro}(a) of the main text.
	The system exhibits multistability for $\Omega/\kappa<1$. The two fixed points $\mathcal{M}_1$ and $\mathcal{M}_2$ have different basins of attraction: 
	depending upon the initial state, the system can exhibit two different phases, namely, a melted and a time-crystal phase.
	For the Dicke subspace, this bistability does not exist, as depicted in Fig.~\ref{fig:intro}(a) of the main text.
	The Dicke subspace corresponds to initial states having $m=1$, which shows that for $\Omega/\kappa<1$, the system always settles down to a saddle fixed point (dashed red line), and for $\Omega/\kappa>1$, it shows oscillatory behavior (solid red line). 
	However, for $m<1$, both time-translation invariant and symmetry-broken phases coexist and are separated by the blue line corresponding to $m=\Omega/\kappa$.
	Initial states having $m > \Omega/\kappa$ on the left of $m=\Omega/\kappa$ (blue) line settle down to the saddle fixed point $\mathcal{M}_2$. 
	The parameter region right to the $m=\Omega/\kappa$ (blue) line corresponds to $m<\Omega/\kappa$ and gives stable oscillations around the fixed point $\mathcal{M}_1$ leading to 
	time-translational symmetry breaking.

	To illustrate the phase-space portrait of the fixed points discussed above, we consider the following coordinate transformation $P=\tan^{-1}\left[ m_y/m_x \right],~Q=m_z$~\cite{iemini2018boundary}. 
	In terms of the new co-ordinates $P$ and $Q$, the fixed points $\{P^*,Q^*\}$ are given as $\mathcal{M}^{P,Q}_1= (\tan^{-1}[\pm 1/\sqrt{(\Omega/m\kappa)^2-1}],0 )$ and $\mathcal{M}^{P,Q}_2=(\pm\pi/2,\pm\sqrt{m^2-(\Omega/\kappa)^2})$. 
	The phase-space portrait of the CTC for different $m$ values is depicted in Fig.~\ref{fig:MF_TimeEvo}(a) for $\Omega/\kappa=0.75$.
	Initial states having $m<\Omega/\kappa$ keep oscillating in the vicinity of the fixed point $\mathcal{M}^{P,Q}_1$. 
	As soon as we choose an initial state with $m>\Omega/\kappa$, the corresponding trajectory in Fig.~\ref{fig:MF_TimeEvo}(a) settles down to the saddle fixed point $\mathcal{M}^{P,Q}_2$. For $\Omega/\kappa>1$, there exists no physical state of the system that can have $m>\Omega/\kappa$ since $m\leq 1$ for the rescaled spin operators. 
	The stability of the oscillations around the fixed point $\mathcal{M}_1^{P,Q}$ can be verified by observing Fig.~\ref{fig:MF_TimeEvo}(a), where each limit cycle trajectory consists of around 4000 cycles of oscillations. These trajectories are obtained by the time evolution of the exact mean-field equations \eqref{eq:MF} (including non-linear terms) and show no deviation even at such long times.
	
	Figure~\ref{fig:MF_TimeEvo}(b) depicts the time evolution of the master equation (Eq.~\eqref{eq:master_BTC} of the main text) with the initial state belonging to the $m=0.1$ subspace, with $\{m_x,m_y,m_z\}=\{0,0,0.1\}$ for $\Omega/\kappa=0.9$. As the system size increases, the oscillation in $m_z$ becomes persistent and approaches the mean-field limit. Any initial state belonging to subspaces with $m>0.9$ will decay and approach a time-invariant steady state irrespective of system size.
	Thus, the exact treatment confirms the mean-field analysis for $\Omega/\kappa<1$. 
	
	\section{Mean-field dynamics of coupled continuous time crystals}
	Here we discuss the mean-field dynamics of the coherently coupled CTCs described by Eq.~\eqref{eq:coupledCTC} of the main text. These equations are formulated in terms of the rescaled spin operators $m^{\alpha}_{\beta} =\langle \hat{S}^{\alpha}_{\beta} \rangle/S$ with $\alpha: \{x,y,z\}$ and $\beta: \{A,B\}$, and are defined as follows, 
	\begin{align}
	\frac{dm^x_{\alpha}}{dt} & = \kappa m^x_{\alpha} m^z_{\alpha} + \frac{\Gamma}{n} m^z_{\alpha} \sum_{\beta\neq\alpha} m^y_{\beta}, \nonumber\\
	\frac{dm^y_{\alpha}}{dt} & = -\Omega_{\alpha} m^z_{\alpha} + \kappa m^y_{\alpha} m^z_{\alpha} - \frac{\Gamma}{n} m^z_{\alpha} \sum_{\beta\neq\alpha} m^x_{\beta},\label{eq:coupledMF} \\
	\frac{dm^z_{\alpha}}{dt} & = \Omega_{\alpha} m^y_{\alpha} - \kappa \left[( m^x_{\alpha})^2 + (m^y_{\alpha})^2 \right] + \frac{\Gamma}{n} \left[ m^y_{\alpha} \sum_{\beta\neq\alpha} m^x_{\beta} - m^x_{\alpha} \sum_{\beta\neq\alpha} m^y_{\beta} \right]. \nonumber
	\label{eq:MF}
	\end{align}
	We numerically solve these coupled differential equations to study the synchronization of coupled CTCs.

	\section{Chimera state and synchronization via seeding}
\begin{figure*}[t!]
		\centering    \includegraphics[width=0.98\textwidth]{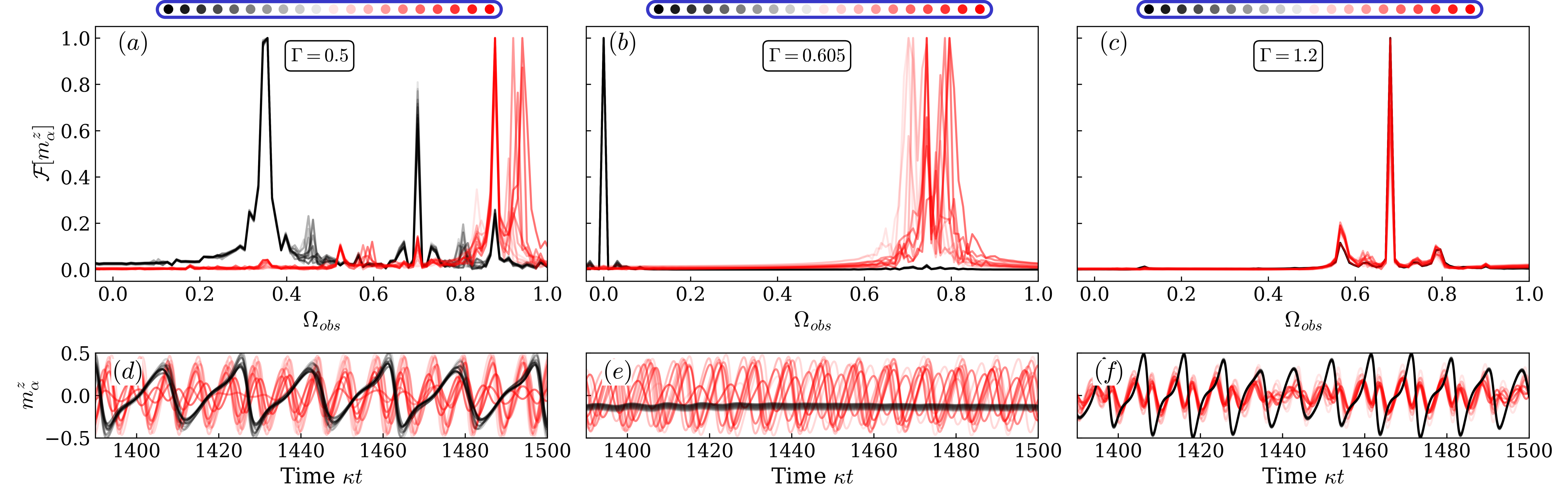}
		\caption{Normalized Fourier transform $\mathcal{F}[m^z_\alpha]$(a-c) and time evolution of $m^z_\alpha$ (d-f) for coupling strength $\Gamma/\kappa=0.5$(a,d), $\Gamma/\kappa=0.605$(b,e) and $\Gamma/\kappa=1.2$(c,f). The frequencies of the uncoupled CTCs are uniformly sampled in the window from $0.2$ to $0.25$ in ensemble $\varepsilon_1$  and from $0.75$ to $0.85$ in ensemble $\varepsilon_2$.} 
		\label{fig:chimera_appendix}
	\end{figure*}

	The initial state leading to the chimera state shown in Fig.~\ref{fig:chimera} of the main text is chosen such that the absolute values of the corresponding Jacobian eigenvalues $\{\vert \Lambda_1^\alpha\vert ; \alpha=1,2,\ldots 20\}$ for the fixed point $\mathcal{M}_1$ are uniformly sampled in the window from $0.2$ to $0.25$ for CTCs in ensemble $\varepsilon_1$ (represented by black color) and from $0.75$ to $0.85$ for CTCs in ensemble $\varepsilon_2$ (represented by red color). For the given initial state, we investigate the effect of increasing the coupling strength between the all-to-all coupled CTCs on the synchronization properties. For coupling strength $\Gamma/\kappa=0.5$, all the CTCs in ensemble $\varepsilon_1$ are synchronized while CTCs in ensemble $\varepsilon_2$ only show partial synchronization, as shown in Figs.~\ref{fig:chimera_appendix}(a,d). Increasing  $\Gamma$ we see more complex dynamics (not shown here), but the system continues to exhibit a chimera state. At $\Gamma/\kappa\approx 0.605$, a transition occurs: the strength of the coupling leads to the melting of the CTCs of ensemble $\varepsilon_1$ as depicted in Figs.~\ref{fig:chimera_appendix}(b,e). Further increasing the coupling strength leads to the synchronization of the CTCs in ensemble $\varepsilon_2$ which seeds oscillations to ensemble $\varepsilon_1$. As a result, at coupling strength $\Gamma/\kappa=1.2$, all the CTCs oscillate with a common frequency as shown in Figs.~\ref{fig:chimera_appendix}(c,f). Such a synchronization via seeding is consistent with what was observed in \cite{seeding2022michal}.

	\section{Cluster synchronization}
\begin{figure*}[htp!]
		\centering    \includegraphics[width=0.98\textwidth]{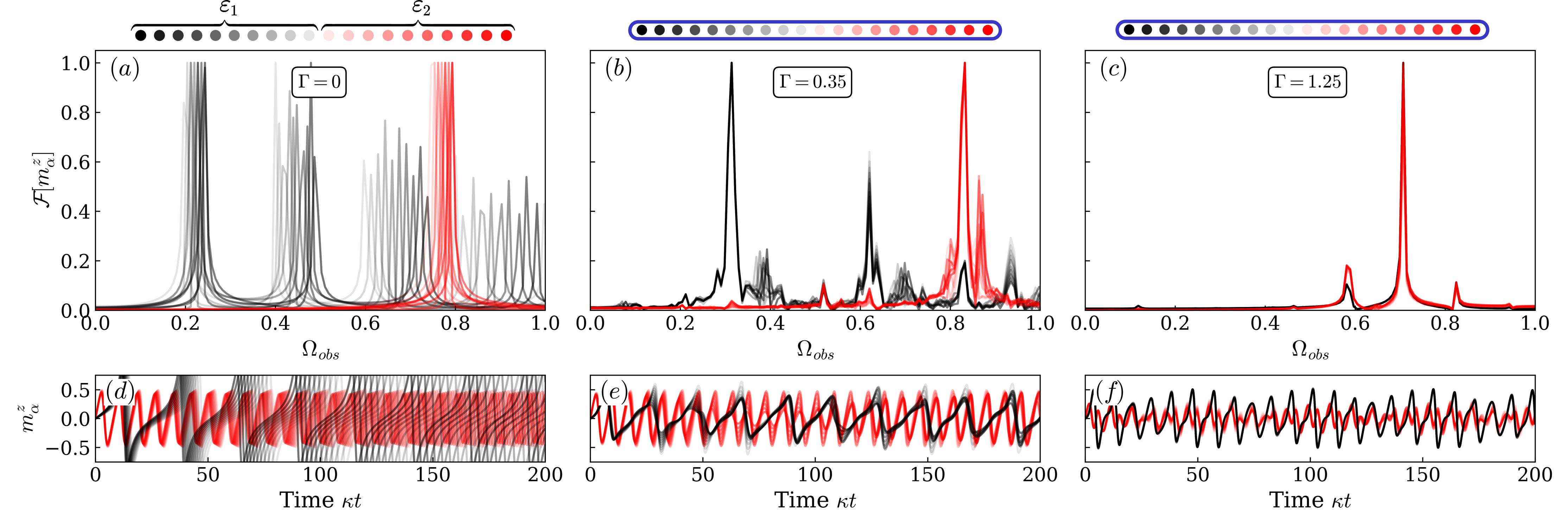}
		\caption{ Here we consider $n=20$ CTCs having the same $\Omega_\alpha/\kappa=0.9$ which are initialized with different $m$ subspaces. Subfigure (a,d) represents uncoupled CTCs. The frequencies of the uncoupled CTCs are uniformly sampled in the window from $0.2$ to $0.25$ in ensemble $\varepsilon_1$  and from $0.75$ to $0.8$ in ensemble $\varepsilon_2$. For finite coupling strength $\Gamma/\kappa=0.35$, there are two clusters of synchronized CTCs with different frequencies, as shown in (b,e). At a coupling strength $\Gamma/\kappa=1.25$, all the CTCs get synchronized as shown in (c,f). } 
		\label{fig:cluster}
	\end{figure*}

	We will now discuss the effect of small changes in the initial state on the synchronization of CTCs. 
	As an example, we consider a change in the initial states for CTCs belonging to ensemble $\varepsilon_2$ such that $\{\vert \Lambda_1^\alpha \vert \}$ are uniformly sampled in the interval from $0.75$ to $0.8$. 
	The choice of initial states for the uncoupled CTCs is shown in Figs.~\ref{fig:cluster}(a,d). Now we study the effect of choosing this distribution on the synchronization of CTCs and we only consider all-to-all coupling in this case. This small change results in the emergence of a new synchronization phase, termed cluster synchronization as shown in Figs.~\ref{fig:cluster}(b,e). As the name suggests, there exist two clusters of synchronized CTCs with different frequencies. Further increasing the coupling strength to $\Gamma/\kappa=1.25$ results in complete synchronization and these two clusters oscillate at the same frequency as shown in Figs.~\ref{fig:cluster}(c,f). These findings demonstrate that different choices of initial states can result in different regimes of synchronization. 
	
	\section{Characterization of different phases using Pearson coefficient and maximum Lyapunov exponent}
	
\begin{figure*}[b!]
		\centering    \includegraphics[width=\textwidth]{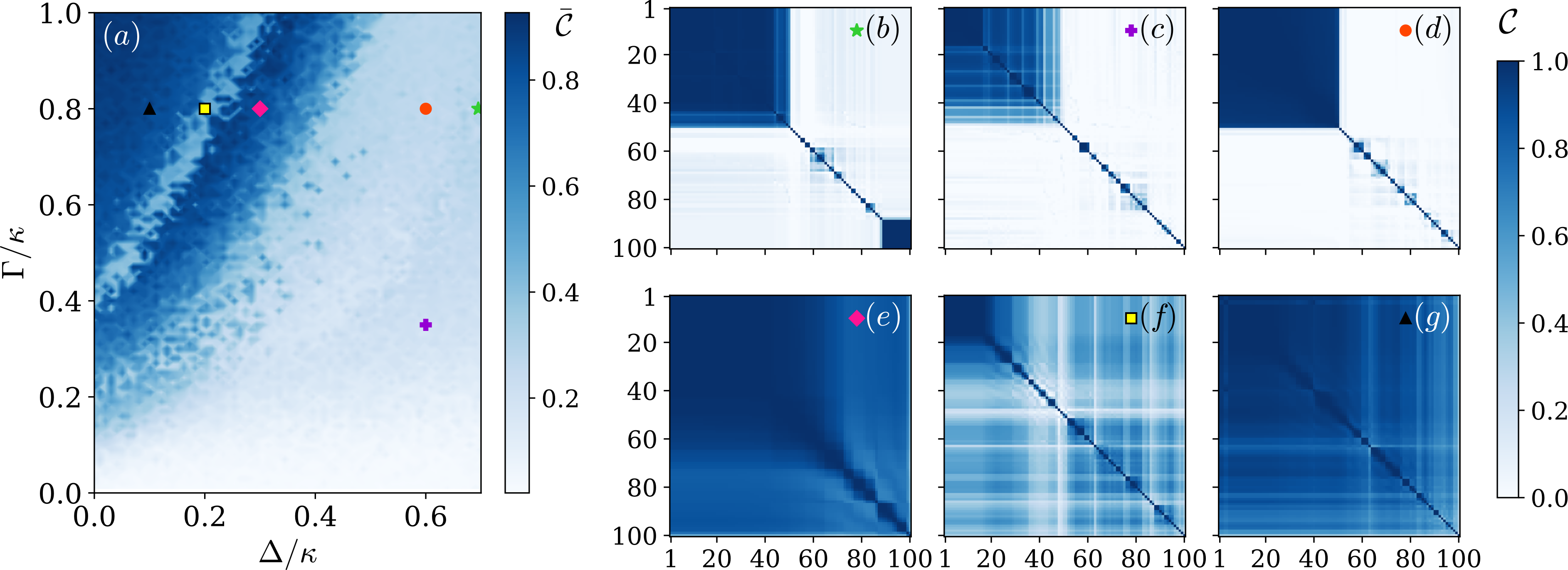}
		\caption{ Ensemble of $n=100$ CTCs having the same $\Omega_\alpha/\kappa=0.9$ that are divided into two subgroups $\varepsilon_{1,2}$ containing $50$ CTCs each. The CTCs are initialized with different $m$ subspaces described by the choice of mean frequencies of the subgroups $\bar{\omega}_{1,2}$ and the standard deviations $\sigma_{1,2}$. (a) Average value of Pearson coefficient $\bar{\mathcal{C}}$.
			(b-g) Pearson coefficient matrices corresponding to different values of $\Gamma$ and $\Delta$ indicated by the symbols (colors) `$\scriptstyle\bigstar$(green)'($\Gamma=0.8,\Delta=0.7$), `\fplus(violet)'($\Gamma=0.35,\Delta=0.6$), `$\bullet$(red)'($\Gamma=0.8,\Delta=0.6$), `$\blackdiam$(pink)'($\Gamma=0.8,\Delta=0.3$), `$\sq$(yellow)'($\Gamma=0.8,\Delta=0.2$)  and `$\blacktriangle$(black)'($\Gamma=0.8,\Delta=0.1$) in subfigure (a). 
		} 
		\label{fig:pearson}
	\end{figure*}

\begin{figure*}[t!]
		\centering    \includegraphics[width=\textwidth]{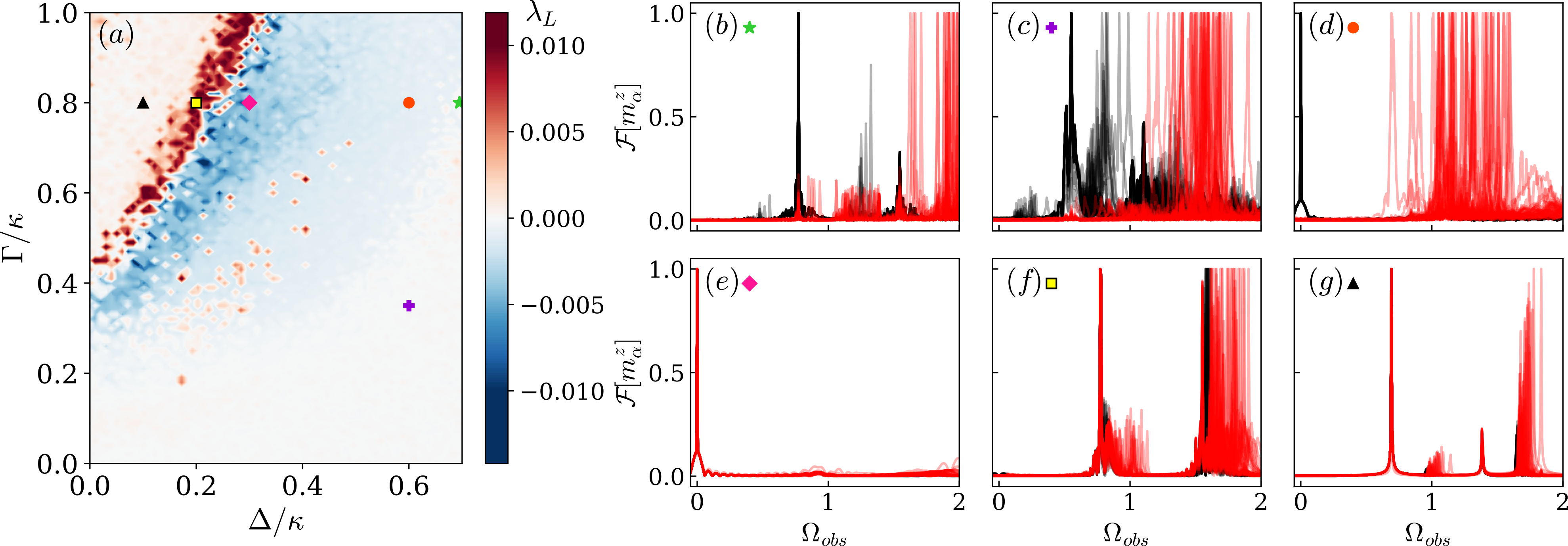}
		\caption{(a) Maximum Lyapunov exponent $\lambda_{L}$ describing the dynamics of coupled CTCs. (b-g) Normalized Fourier transform $\mathcal{F}[m^z_\alpha]$ of the time evolution of $m^z_\alpha$ for different values of $\Gamma$ and $\Delta$ indicated by the symbols (colors) `$\scriptstyle\bigstar$(green)'($\Gamma=0.8,\Delta=0.7$), `\fplus(violet)'($\Gamma=0.35,\Delta=0.6$), `$\bullet$(red)'($\Gamma=0.8,\Delta=0.6$), `$\blackdiam$(pink)'($\Gamma=0.8,\Delta=0.3$), `$\sq$(yellow)'($\Gamma=0.8,\Delta=0.2$)  and `$\blacktriangle$(black)'($\Gamma=0.8,\Delta=0.1$) in subfigure (a). All the other parameters are as in Fig.~\ref{fig:pearson}. } 
		\label{fig:lyapunov}
	\end{figure*}

	Here, we provide a detailed characterization of the various synchronization regimes discussed in the main text.
	We consider an ensemble of $n=100$ CTCs that are divided into two subgroups $\varepsilon_{1,2}$ containing $50$ CTCs each.
	The frequencies of the CTCs in each subgroup $\varepsilon_{1,2}$ are sampled from Gaussian distributions with distinct mean frequencies $\bar{\omega}_{1,2}$ and standard deviations $\sigma_{1,2}$.
	As in Fig.~4(e,f) of the main text, we use $\sigma_{1}=\sigma_{2}=0.1$, and vary $\Delta=\bar{\omega}_{1}-\bar{\omega}_{2}$ and $\Gamma$ to explore six regimes of synchronization marked by different symbols in Figs.~\ref{fig:pearson} and \ref{fig:lyapunov}.
	We study the average value of the Pearson coefficient $\bar{\mathcal{C}}$ and the maximum Lyapunov exponent $\lambda_L$ defined in the main text to characterize these regimes.
	
	\bigskip
	
	\paragraph{Partial Synchronization ---} We first consider the three different cases marked by the symbols `$\scriptstyle\bigstar$'($\Gamma=0.8,\Delta=0.7$), `\fplus'($\Gamma=0.35,\Delta=0.6$) and `$\bullet$'($\Gamma=0.8,\Delta=0.6$) in Figs.~\ref{fig:pearson} and \ref{fig:lyapunov}.
	These three cases exhibit $\bar{\mathcal{C}}<1$ as depicted in Fig.~\ref{fig:pearson}(a).
	However, they possess distinct dynamical properties, as evidenced by the maximum Lyapunov exponent shown in Fig.~\ref{fig:lyapunov}(a).
	Therefore, we examine the Pearson correlation matrix $\mathcal{C}$ along with the Lyapunov exponent to distinguish between the following different cases of partial synchronization:
	
	\begin{itemize}
		\item \textbf{Coexistence of chimera and cluster synchronization:} For $\Gamma=0.8$ and $\Delta=0.7$ (denoted by `$\scriptstyle\bigstar$'), the CTCs are in a stable limit-cycle state since $\lambda_L= 0$ (as shown in Fig.~\ref{fig:lyapunov}(a)). 
		The correlation matrix depicted in Fig.~\ref{fig:pearson}(b) reveals two prominent clusters of synchronized CTCs, while the remainder of the matrix is diagonal, indicating unsynchronized CTCs.
		A similar behavior can be observed in Fig.~\ref{fig:lyapunov}(b), where two synchronized subsets of CTCs have equal frequencies, while the rest of the unsynchronized CTCs oscillate with different frequencies.
		Thus, cluster synchronization and a chimera state can coexist.
		
		\item \textbf{Chimera state:} For $\Gamma=0.35$ and $\Delta=0.6$ (denoted by `\fplus'), there is only one prominent cluster of synchronized CTCs while the other CTCs remain unsynchronized.
		This can be understood from Fig.~\ref{fig:pearson}(c), where the CTCs of subgroup $\varepsilon_1$ show a high amount of correlation while the remainder of the CTCs are almost uncorrelated.
		Therefore, the CTCs of subgroup $\varepsilon_1$ have a common frequency, while the other CTCs oscillate with different frequencies as shown in Fig.~\ref{fig:lyapunov}(c).
		Thus, the system exhibits a chimera state. 
		
		\item \textbf{Chimera state with partial oscillation death:} We next discuss a case in which the Pearson correlation matrix looks qualitatively similar to that of a chimera state (see Fig.~\ref{fig:pearson}(c)). This can be observed for $\Gamma=0.8,\Delta=0.6$ (denoted by `$\bullet$'), as shown in Fig.~\ref{fig:pearson}(d).
		Note that $\lambda_L<0$, see Fig.~\ref{fig:lyapunov}(a).
		Thus, some of the CTCs cease to oscillate leading to partial oscillation death.
		The same conclusion follows from Fig.~\ref{fig:lyapunov}(d) where all the CTCs of subgroup $\varepsilon_1$ have zero frequency and hence melt down.
		The steady state of the melted CTCs corresponds to highly correlated fixed points, and the correlation matrix exhibits a block diagonal structure similar to what was observed for chimera states.
	\end{itemize}
	
	As we have seen, $\bar{\mathcal{C}}$ alone does not provide sufficient information to distinguish different cases of partial synchronizations.
	Therefore, we investigated the correlation matrix and the maximum Lyapunov exponent to differentiate between these cases. 
	
	\bigskip

	\paragraph{Oscillation death ---} We now discuss the case of complete oscillation death, where all the CTCs melt down and stop oscillating. 
	This can be observed for $\Gamma=0.8$ and $\Delta=0.3$ (represented by `$\blackdiam$'), where $\lambda_L\ll0$, as shown in Fig.~\ref{fig:lyapunov}(a).
	Indeed, from Fig.~\ref{fig:lyapunov}(e), we can observe that all the CTCs have zero frequency.
	Therefore the system settles down to a fixed point corresponding to complete oscillation death.
	The Pearson coefficient matrix shows a large amount of correlation between all the CTCs.
	\bigskip
	
	\paragraph{Mean-field chaos ---} By further reducing the detuning to $\Delta=0.2$ while keeping $\Gamma=0.8$ (represented by `$\sq$'), the mean-field dynamics becomes chaotic, indicated by a positive value of $\lambda_L$ in Fig.~\ref{fig:lyapunov}(a).
	This leads to a lower value of $\bar{\mathcal{C}}$ as shown in Fig.~\ref{fig:pearson}(a). 
	Although the correlation matrix exhibits a reduced amount of correlations between the CTCs due to the chaotic nature of the mean-field dynamics,
	the finite value of its off-diagonal elements indicates the presence of chaotic synchronization.
	\bigskip
	
	\paragraph{Complete synchronization ---} Finally, for small detuning $\Delta=0.1$ and $\Gamma=0.8$ (represented by `$\blacktriangle$'), the system becomes completely synchronized such that $\bar{\mathcal{C}}\approx 1$ and $\lambda_L\approx 0$.
	This regime is indicated by Fig.~\ref{fig:pearson}(g), where all the off-diagonal elements of $\mathcal{C}$ are non-zero, representing a large amount of correlation between all CTCs.
	The complete synchronization can also be observed from Fig.~\ref{fig:lyapunov}(g), where all CTCs oscillate with a common frequency.

\end{document}